# Active microcantilevers based on piezoresistive ferromagnetic thin films

Harish Bhaskaran[1], Mo Li[1], Daniel Garcia-Sanchez[1], Peng Zhao[2], Ichiro Takeuchi[2] and Hong X. Tang[1][*]

[1]*Department of Electrical Engineering, Yale University, New Haven CT 06511*
[2]*Department of Materials Science and Engineering, University of Maryland, College Park, MD 20742*

**Abstract:**

We report the piezoresisitivity in magnetic thin films of $Fe_{0.7}Ga_{0.3}$ and their use for fabricating self transducing microcantilevers. The actuation occurs as a consequence of both the ferromagnetic and magnetostrictive property of $Fe_{0.7}Ga_{0.3}$ thin films, while the deflection readout is achieved by exploiting the piezoresisitivity of these films. This self-sensing, self-actuating micromechanical system involves a very simple bilayer structure, which eliminates the need for the more complex piezoelectric stack that is commonly used in active cantilevers. Thus, it potentially opens opportunities for remotely actuated, cantilever-based sensors.

---

[*] Corresponding author (hong.tang@yale.edu)

Microcantilevers have been at the heart of atomic force microscopy (AFM) and a range of chemical and biological sensing applications. The transduction of microcantilevers often involves off-chip actuators such as a piezo shaker, as well as off-chip displacement readout, such as optical deflection. Active cantilevers that contain integrated actuation and sensing elements do not require alignment of actuating and sensing components. Therefore they offer a higher level of system integration and are more suitable for portable applications.

Active cantilevers can be constructed by integrating piezoelectric[1-3] or magnetic thin films[4,5] to exert driving forces. Using magnetic forces to actuate micromechanical devices is advantageous for applications that are susceptible to perturbations, as magnetic fields can be generated remotely. To build active magnetic cantilevers, on-chip readout of the mechanical motion is also desired. One scheme is to integrate a layer of piezoelectric material so that a potential difference is generated in the piezoelectric layer when the device is stressed by magnetic force. This so-called magnetoelectric effect (ME) has been utilized in sensors using a variety of magnetostrictive materials[6-9]. However, when devices are scaled to smaller dimensions, the use of thick multilayer structural stack that is necessary for harvesting ME effects is often undesirable. In this letter, instead of using the ME effect, we utilize the intrinsic piezoresistivity in a thin film $Fe_{0.7}Ga_{0.3}$ [7] to detect and magnetically actuate micromechanical motion. This method has significant advantages over the ME effect in that it permits further scaling down of device dimensions meanwhile eliminating the need for integration of piezoelectric materials and electrodes which involves extensive processing.

Although piezoresistivity in metals is a well-known phenomenon, this property is rarely applied as transducing elements in microdevices due to the nominally low gauge factor $\gamma$ (defined as $(\delta R/R)/(\delta L/L)$, where $L$ and $R$ are the length and resistance of a strain gauge respectively). Recently, gold thin films have been shown to be efficient piezoresistive sensors on nanoscale mechanical devices[10,11]. In this letter, we demonstrate the piezoresistivity in metallic $Fe_{0.7}Ga_{0.3}$ thin films and its use for both magnetic actuation and piezoresistive sensing in micromechanical devices. Our fabrication process starts with micro-machined membranes of high-stress silicon nitride (330 nm thick) on a silicon wafer made by anisotropic KOH etch. A thin layer of $Fe_{0.7}Ga_{0.3}$ (100 nm) is sputtered on the membranes (for details of the material, see Zhao et al[7]). The cantilever structures are then patterned using photolithography. Ion milling is used to etch the $Fe_{0.7}Ga_{0.3}$ layer. This is followed by etching the $Si_3N_4$ in a reactive ion etcher using $CHF_3$ to pattern and release the cantilever beams. The cantilevers are designed to have a stress concentration region at the base in order to optimize the piezoresistive response (Fig. 1a). Based upon finite element simulations of stress distribution in the cantilever, we estimate that the stress concentration region accounts for ~85% of the measured resistance in the fabricated devices, thus contributing to the bulk of the piezoresistive response. An SEM image of a fabricated cantilever of dimensions 100 μm (length) × 20 μm (width) is shown in Fig. 1(a). We note that due to the intrinsic stress in the deposited $Fe_{0.7}Ga_{0.3}$ layer, bending of the cantilever structures is observed. This bending causes imperfect alignment of magnetic field with the cantilever beam. As a result, the magnetic actuation is a combined effect of magnetic torque and magnetostriction of $Fe_{0.7}Ga_{0.3}$ thin films. As we demonstrate later, a qualitative comparison of the relative magnitude of each of these effects can be obtained.

We first investigate the piezoresistivity of the $Fe_{0.7}Ga_{0.3}$ film by directly driving a piezo disk mounted below the sample holder. The nominal resistance of the cantilevers is around 1.7 kΩ. All measurements are carried out in high vacuum at the resonance frequency of the cantilever in the dynamic regime. A schematic for the piezoresistive measurement is shown in Fig. 1(b). The device resistance $R_{cant}$ changes as a function of strain induced by the bending of the cantilever. A DC bias voltage $V_{cant}$= 1V is supplied to the cantilever to convert the resistance variation to voltage signal, which is amplified with a pre-amp. The frequency response of the cantilever measured with a network analyzer is shown in Fig. 1(c). $Fe_{0.7}Ga_{0.3}$ cantilevers are observed to resonate at their fundamental modes in the frequency range of 16-24 kHz with quality factors ranging from 3000-6000 in high vacuum (<$1\times10^{-4}$ Torr). This compares well with the expected resonance frequency of 21 kHz for such cantilevers, as computed using finite element methods (modeled as a no-stress silicon-nitride cantilever with $Fe_{0.7}Ga_{0.3}$ film on top). We compare the piezoresistive effect of the $Fe_{0.7}Ga_{0.3}$ film with that in a gold film of the same thickness of 100 nm, as shown in Fig. 1(c). Accounting for frequency differences between the two cantilevers (explained by the difference in density and young's modulus of Au and $Fe_{0.7}Ga_{0.3}$) and biasing conditions, we conclude that the gauge factor of $Fe_{0.7}Ga_{0.3}$ thin films is only ~50% of the gauge factor of gold films of the similar thickness.

In a second experiment, we remotely actuate the cantilevers using an external magnetic field, without the use of a piezo actuator. Two cantilevers, one parallel to and the other perpendicular to the applied magnetic field are excited using an AC magnetic field with RMS amplitude of ~0.5 Oe. As shown in Fig. 2, the fundamental mode of the cantilevers can be driven using a magnetic field for both cantilevers, whereas the piezoresistive cantilevers made from gold films fail to show such a response. Thus, this demonstrates that the cantilevers with $Fe_{0.7}Ga_{0.3}$ films can be self-sensing and be actuated using remote magnetic fields. The difference in the resonance frequencies of the two cantilevers is due to fabrication variation.

In order to ascertain the mechanism of actuation, we note that the magnetostrictive effect is different from a simple torque effect. In thin sputtered films of nanocrystalline materials (as is the present case), the film is multi-domain with no dominant magnetic easy axis. Thus, a magnetic field applied in the plane of the cantilever and in a direction perpendicular to the cantilever's length (in-plane transverse drive) would produce a pure magnetostrictive effect, whereas the torque effect is absent since the film magnetization is always aligned with the field[12]. For the magnetic field applied parallel to the cantilever length (longitudinal drive), although every effort is made in this experiment to keep the cantilevers length well aligned to the magnetic fields, there is an experimental uncertainty as to the presence of residual misalignment. This misalignment introduces additional magnetic torque actuation of the cantilever. As a result, the observed cantilever vibration amplitude is higher with longitudinal drive than with transverse drive. To isolate the relative effect of each and directly access the magnetoelastic stress effect, we monitor the resonance frequency shift of the cantilever when subjected to DC magnetic fields in two normal in-plane directions. The magnetic torque effect is expected to be present only when the field is applied with the longitudinal drive configuration especially for cantilevers that are bent due to intrinsic stress, whereas it would be significantly smaller with the transverse drive configuration (inset in Fig. 3 (a)).

The frequency response of the cantilevers is monitored with a phase lock loop while driving the cantilever using the piezo disk. The frequency variation of one cantilever is shown in Fig. 3 (a), whereby the transverse DC magnetic field (H) is swept from -270 Oe to +270 Oe. As the cantilever is now driven by the piezo disk with constant power, the vibration amplitude remains nominally the same, *i.e.* there is no amplitude modulation due to applied magnetic fields. The frequency is observed to follow a hysteresis curve similar to that expected in magnetic susceptibility of ferromagnetic materials, with a saturation field of ~120 Oe. This saturation behavior rules out the contribution of magnetic torque effect as the torque ($L = M \times B$) is proportional to the biasing field and thus will not saturate at high magnetic field. On the contrary, magnetostrictive stress along the length due to a transverse magnetic field can be written as[13] $S = -2K\lambda E_{FeGa}(1+v_{FeGa})$, where $K$ is a constant that adjusts for non-directionality of the sputtered nanocrystals, $\lambda$ is the coefficient of magnetostriction, $E_{FeGa}$ is the Young's modulus and $v_{FeGa}$ is the Poisson's ratio of $Fe_{0.7}Ga_{0.3}$. The effective spring constant for the bi-layered cantilever can be expressed as $k_{eff} = E_{eff}G_k$, where $G_k$, is the geometrical constant, which depends on the cantilever geometry and $E_{eff}$ is the effective Young's modulus[12]. The effective spring constant is a function of the Young's modulus of the $Fe_{0.7}Ga_{0.3}$ film as well as that of the underlying silicon nitride. Given that the magnetostriction is only a property of $Fe_{0.7}Ga_{0.3}$, this implies that the $E_{eff}$ is modulated only due to a magnetostrictive effect in the $Fe_{0.7}Ga_{0.3}$ layer. The frequency $\omega_0$ is then given by $\omega_0 = (k_{eff}/m)^{-1/2}$, where $m$ is the cantilever modal mass. Thus, a change in $E_{FeGa}$ due to $S$ would result in a change in $E_{eff}$ and corresponding variation in the cantilever frequency. Thus, it follows that our observed frequency change hysteresis provides evidence that the effect is predominantly magnetostrictive.

We repeat the measurements shown in Fig. 3a, but reconfigure our setup to drive the cantilevers remotely with an ac magnetic field (as opposed to driving the cantilever using the piezo disk as in the previous case), on two orthogonally oriented cantilevers (cantilevers 2 and 3) whose resonances closely match Fig. 2. The ac magnetic field is co-aligned with the DC bias field. The frequency variation for both sets is shown in Fig. 3b and c. As shown in Fig. 3b, the cantilever 2 that is driven using a magnetic field transverse to its length exhibits a behavior similar to Fig. 3a of Cantilever 1, albeit noisier, due to the reduced vibration amplitudes. However, for the cantilever 3 driven with longitudinal magnetic fields, the frequency shows no saturation for higher values of H (Fig. 3c). This indicates that the effect is mostly a torque effect in this configuration. An analysis similar to the one described in Ref 13 shows that, in this case, because of the cantilever's orientation, there exists a component of the magnetic field in the direction perpendicular to the cantilever plane, and the torque effect would modify the frequency in the observed manner. Furthermore, as the effect is more pronounced in this case, this gives a qualitative indication that the torque effect is a greater effect in this configuration. We expect that further device optimization, in particular stress compensation during growth process, could alleviate the cantilever bending and thus allow us to fully decouple the magnetostrictive effect and torque effect in our device system.

The results here indicate that $Fe_{0.7}Ga_{0.3}$, a magnetostrictive material, can be deposited on the surface of MEMS and NEMS devices as sensing and actuation elements. Such a mechanical resonator can be actuated by a remote magnetic field. At the same time, self sensing is achieved by exploiting the intrinsic piezoresistivity of $Fe_{0.7}Ga_{0.3}$. Although these experiments suggest that the transduction is a combination of magnetostriction and torque effects, careful design of mechanical resonators can exploit only one effect or the other in future. The total thickness of the sensor-actuator layer is 100 nm, compared to several microns in devices that use the ME effect, thus significantly shrinking device dimensions. Such a combination of two effects in the same material would find use in applications that require low complexity, high robustness and reduced dimensions.

The authors acknowledge funding from DARPA/DSO's HUMS project under grant number FA8650-09-C-1-7944.

**Figures:**

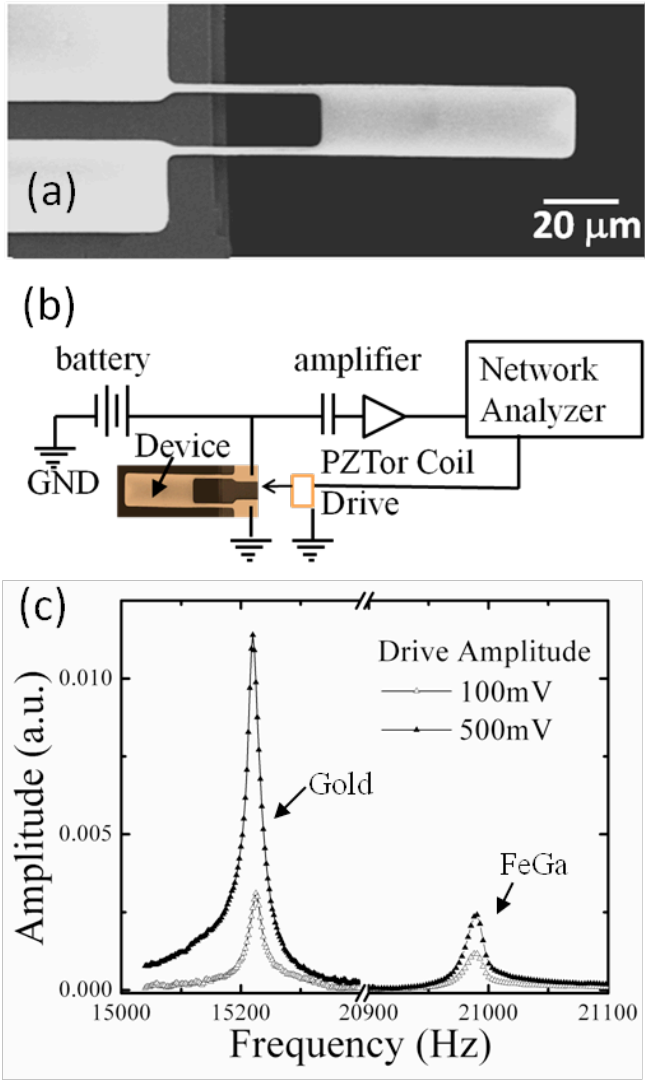

FIG. 1 (a) Scanning electron micrograph of a fabricated $Fe_{0.7}Ga_{0.3}$ microcantilever. Bright areas indicate $Fe_{0.7}Ga_{0.3}$ on Silicon Nitride (b) Schematic of circuit used for the measurement. A piezo disk shaker or ac coils are used to drive the cantilever and (c) Piezoresisitive read-out of resonances of a gold (left) and a $Fe_{0.7}Ga_{0.3}$ microcantilever (right) of the same geometry, showing the relative amplitude of resonance peaks for a measure of relative piezoresistive sensing efficiency. Responses at two drive amplitudes (to the piezoelectric disc drive) are displayed.

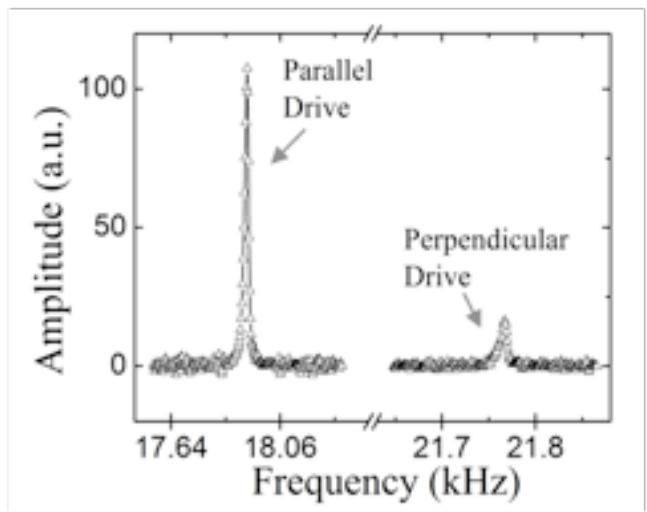

FIG. 2 Frequency response due to magnetically driven actuation of two different cantilevers, one aligned in parallel (longitudinal) to the applied field (left), while the other along the orthogonal (transverse) directions (right).

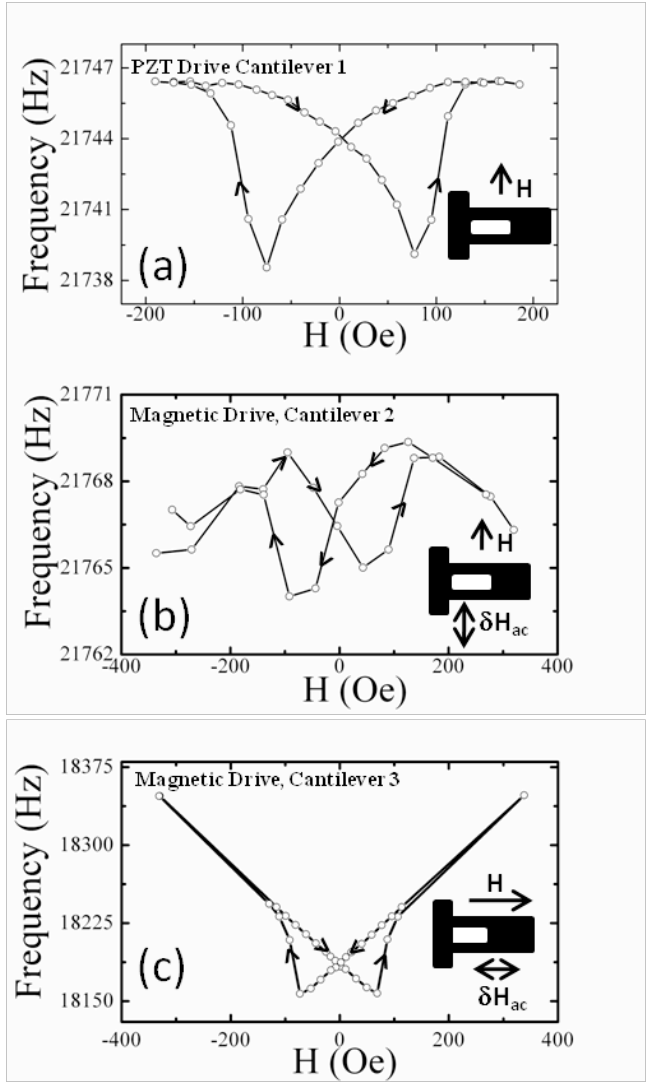

FIG. 3 (a) Variation of resonant frequency of an $Fe_{0.7}Ga_{0.3}$ microcantilever (cantilever 1) driven with a piezo shaker, and subject to a varying magnetic field ($H$) along an in-plane direction perpendicular to the length of the cantilever (inset) and (b) Frequency variation of magnetically actuated cantilevers with transverse fields (cantilever 2, directions indicated in inset) and (c) parallel (cantilever 3). Arrows on data indicate direction in which dc magnetic field $H$ is swept.